\documentclass{aa}
\usepackage{graphicx}
\usepackage{txfonts}
\usepackage{natbib}
\usepackage[bookmarksopen=false,bookmarksnumbered=true,breaklinks=true,colorlinks=true,linkcolor=blue,citecolor=blue]{hyperref}

\usepackage{multicol}
\usepackage{soul} 
\setstcolor{red}
\newdimen\captwidth
\newdimen\figwidth
\newcommand{\mjup}{\ensuremath{\textnormal{M}_{\textnormal{Jup}}}}
\newcommand{\msun}{\ensuremath{\textnormal{M}_{\odot}}}

\newcommand{\excs}{\extracolsep{\fill}}
\newcommand{\bp}{$\beta$~Pic}
\newcommand{\bpb}{$\beta$~Pic~b}
\newcommand{\bpc}{$\beta$~Pic~c}
\newcommand{\bpd}{$\beta$~Pic~d}
\newcommand{\bpw}{$\beta$~Pictoris}
\usepackage{soul}
\usepackage{xcolor}

\newcommand\T{\rule{0pt}{2.6ex}}       
\newcommand\B{\rule[-1.2ex]{0pt}{0pt}} 

\begin{document}
\title{Dynamics of the $\beta$ Pictoris planetary system and possibility of an additional planet}

\author{A. Lacquement\inst{1} \and  H. Beust\inst{1} \and V. Faramaz-Gorka\inst{2} \and G. Duch\^ene\inst{1,3}}

\institute{
$^{1}$Univ. Grenoble Alpes, CNRS, IPAG, 38000 Grenoble, France\\
$^2$Department of Astronomy and Steward Observatory, University of Arizona, 933 N Cherry Ave., Tucson, AZ 85719, USA\\
$^3$Department of Astronomy, University of California, Berkeley, CA 94720, USA}
\date{Received October 16, 2024 / Accepted December 27, 2024}  \offprints{A. Lacquement}
\mail{antoine.lacquement@univ-grenoble-alpes.fr}
\titlerunning{Dynamics of the \bpw\ planetary system}
\authorrunning{A. Lacquement et al.}
\abstract
{The \bpw\ system is a well-known young planetary system, extensively studied for more than 40 years. It is characterized by a dusty debris disk, in addition to the presence of two already known planets. This makes it a particularly interesting case for studying the formation and evolution of planetary systems at a stage where giant planets have already formed, most of the protoplanetary gas has dissipated, and terrestrial planets could emerge.}
{Our goal here is to explore the possibility of additional planets orbiting beyond the outermost known one, \bpw~b. More specifically, we aim to assess whether additional planets in the system could explain the discrepancy between the predicted cutoff of the disk inner cavity at $\sim$28~au with only two planets, and the observed one at $\sim$50~au.}
{We performed an exhaustive dynamical modeling of the debris disk and the carving of its inner edge, by introducing one or two additional planets beyond \bpw~b, coplanar with the disk. Guided by theoretical predictions for the parameter space -- mass, semi-major axis, eccentricity -- allowed for additional planets, we further carried out a set of N-body simulations, using the symplectic integrator RMVS3.}
{Our simulations indicate that an additional planet with a low eccentricity of 0.05, a mass between 0.15 and 1~\mjup, and a semi-major axis between 30 and 36~au would be consistent with the observations of an inner debris disk edge at 50~au. We also explored the hypotheses of a higher eccentricity and the presence of two additional lower-mass planets instead of one, which could also account for these observations.} 
{While we find that one or even two additional planets could explain the observed location of the disk inner edge, these hypothetical planets remain in most cases below the current observational limits of high-contrast imaging. Future observational campaigns with improved sensitivity will help to lower these limits and perhaps detect that planet.}

\maketitle
\section{Introduction}
\label{Introduction}

The star \bpw\ (hereafter \bp) is a young ($18.5^{+2.0}_{-2.4}$~Myr, \citeauthor{Miret2020} \citeyear{Miret2020}) and nearby ($19.63\pm0.06$~pc, \citeauthor{Gaia2020} \citeyear{Gaia2020}) southern star that has consistently captured the attention of astronomers since the discovery of its circumstellar dust disk by \citet{Smith1984}, the first of its kind ever imaged. Earlier that same year, the presence of circumstellar dust was also inferred for the first time through infrared observations of Vega \citep{Aumann1984}. Since these groundbreaking discoveries, it has been established that at least $\sim$20\% of FGK stars host such a circumstellar dust disk \citep{Eiroa2013,Sibthorpe2018}.

Initially, the observed dust was believed to be of primitive origin. However, \citet{Backman1993} demonstrate that this dust is too short-lived for that paradigm to be valid. Indeed, the lifetime of dust grains, when considering removal mechanisms such as radiation pressure, collisions, and Poynting-Robertson drag \citep[see, e.g., the review by][]{Krivov2010}, is shorter by orders of magnitude than the ages of the stars around which they are detected. Consequently, this dust is understood to be second-generation material, widely believed to be continuously replenished by an underlying population of kilometer-sized bodies. These planetesimals serve as a reservoir, capable of sustaining dust production throughout the lifetime of stars, either through slow evaporation \citep{Lecavelier1996} or collisions \citep{Backman1993,Artymowicz1997}.

By observing the dust and understanding the mechanisms behind particle production and movement, it is possible to deduce the structure of the planetesimal reservoir from which the dust originates. However, this connection is rarely straightforward. Radiation pressure causes small dust particles to deviate from the orbits of their parent bodies, often reaching apoastrons far from their initial production sites \citep{Lecavelier1996}. Some of these particles are even observed in scattered light at significant distances from \bp\ \citep[see, e.g.,][]{Janson2021}. To address this complexity in the \bp's system, models have been developed to constrain the distribution of parent bodies based on the observed dust. Several authors have made observations and/or proposed models suggesting that the inner edge of the disk is at $\sim$50~au \citep{Augereau2001, Dent2014, Apai2015, Ballering2016}. However, the edge-on orientation of \bp's disk causes irregularities at its inner edge in these models due to projection effects, introducing significant uncertainty into the estimates. This problem also occurs at millimetre wavelengths. Millimeter emission is a valuable tool for directly revealing the structure of planetesimal disks. It is dominated by large dust particles, which are only minimally affected by radiation pressure, follow the same orbits as their parent bodies, and remain close to them. However, excellent resolution at these wavelengths is crucial for obtaining accurate information about the inner edge of the disk in this edge-on configuration. For example, \citet{Matra2019} reports observations with at least twice the resolution of \citet{Dent2014}, confirming an inner edge of the disk at $\sim$50~au.

The debris disk around \bp\ is one of the most extensively studied, offering valuable insights into the dynamical and evolutionary processes of disks interacting with planets. To date, at least one giant planet has been hypothesized to explain various phenomena observed in the disk. For example, the misalignment between its inner and outer regions \citep{Mouillet1997}, the asymmetries observed between the two arms of the disk \citep{Kalas-Jewitt1995, Heap2000}, and the recurring detection of exocomets in the star's spectrum \citep{Beust-Morbidelli2000}, all suggest the gravitational influence of at least one planet \citep{Augereau2001}. Indeed, a planet was observed. \bpb, a gas giant with a mass of $\sim$12~\mjup, was detected using high-contrast imaging techniques \citep{Lagrange2009}. Its orbit was since refined through regular monitoring, revealing a moderately eccentric orbit with an eccentricity of $\sim$0.1 at a distance of $\sim$10~au from the star \citep{Lacour2021}.

Since then, a number of intriguing features have been identified in the disk, potentially linked to dynamical gravitational interactions with planets. Mid-infrared observations reveal a large dust clump on the western side of the disk \citep{Telesco2005}. Submillimeter observations by Atacama Large Millimeter/submillimeter Array (ALMA) also detect a concentration of CO coinciding spatially with this dust clump \citep{Dent2014, Matra2017}. However, its nature remains uncertain. If the clump is in motion, it could indicate the presence of a gas vortex, potentially caused by an unseen planet trapping the dust \citep{Skaf2023}. Conversely, a stationary clump might result from a past massive collision \citep{Han2023}, possibly consistent with the "cat's tail" recently observed by James Webb Space Telescope (JWST), which may represent debris from a collision between two large planetesimals \citep{Rebollido2024}. It should be noted that these features are located much further out in the disk than the region studied in this work, although there may be a link with the dynamical gravitational interactions of planets closer to the star.

Furthermore, a second planet, another gas giant with a mass of $\sim$9~\mjup, \bpc, was detected. It was discovered through radial velocity monitoring \citep{Lagrange2019} and subsequently confirmed by high-contrast imaging \citep{Nowak2020}. This planet, while slightly less massive than \bpb, orbits closer to the star at $\sim$3~au, following a more eccentric orbit with an eccentricity of $\sim$0.3 \citep{Lacour2021}.

Our goal in this paper is to establish a connection between the \bp's planetary system and the structure of the planetesimal disk, particularly the inner cavity at $\sim$50~au. We assumed this cavity is shaped by the gravitational influence of the planets, neglecting other dynamical effects that remain poorly constrained today. We adopted an age of 20~Myr for the \bp\ system and assumed that the inner edge of the disk lies precisely at 50~au. However, it is important to recognise the uncertainties associated with this model, as the results can be refined with the arrival of new, more precise observations. In Sect.~\ref{2KnownPlanets}, we first conduct a numerical exploration of the gravitational effects of the two known planets, concluding that they cannot sculpt the disk up to 50~au. In Sect.~\ref{1AdditionalPlanet}, we explore the possibility of an additional, yet-to-be-discovered planet orbiting outside \bpb\ that could fulfill this role, and we constrain the characteristics of this hypothetical planet. In Sect.~\ref{AlternativeScenarios}, we consider alternative scenarios, such as a more eccentric planet or the presence of two additional planets instead of one. We present our conclusions in Sect.~\ref{Conclusion}.

\section{Exploration with the two known planets}
\label{2KnownPlanets}

Here, we focus on the dynamical effects of the two known planets in the \bp\ system on the debris disk's inner edge and demonstrate that they are insufficient to create this edge at 50~au.

\subsection{Semi-analytical theory}
\label{Theory}
\begin{figure*}
\makebox[\textwidth]{
\includegraphics[width=0.3333\textwidth]{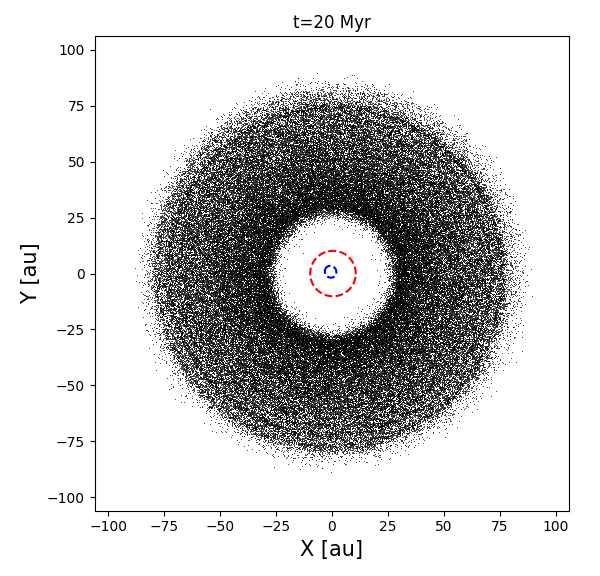} \hfil
\includegraphics[width=0.3333\textwidth]{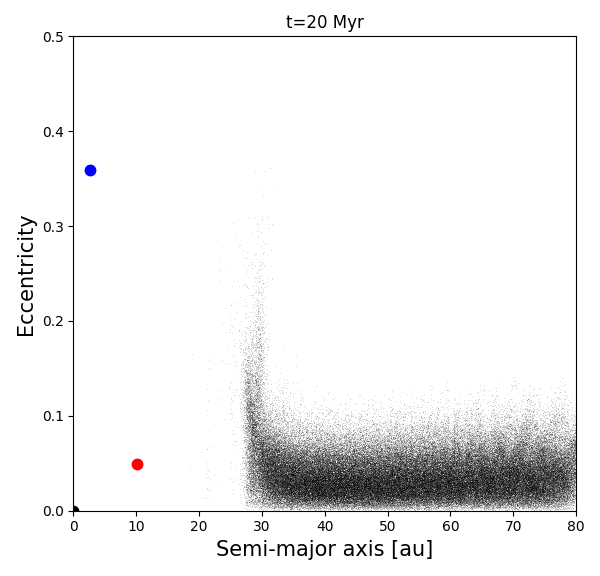} \hfil
\includegraphics[width=0.3333\textwidth]{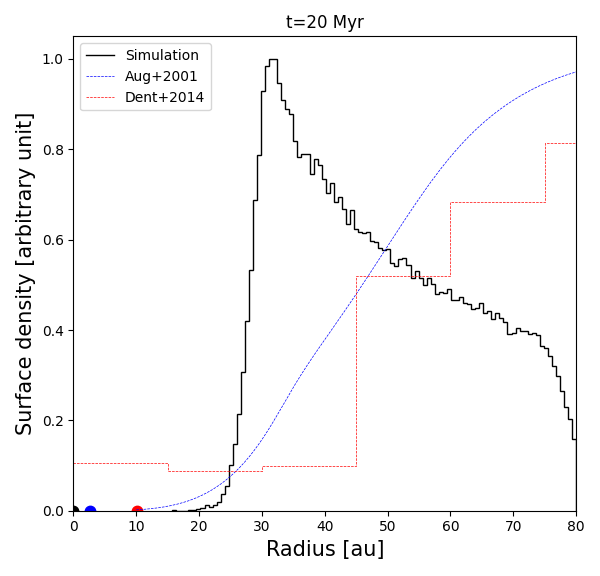}}
\caption{Preliminary simulation of the dynamics of the \bp\ planetary system with only the two planets known to date, \bpb\ (red) and \bpc\ (blue), and the disk of planetesimals (black). The initial orbital parameters of \bpb\ and c are taken from Table~\ref{tab:bcparameters}. \emph{Left:} upper view of the system. The planetesimals are depicted by small black dots, and the orbits of the planets by colored dashed lines. \emph{Middle:} view of the system in terms of semi-major axis and eccentricity. The planets are here represented by colored points. \emph{Right:} radial profile of the planetesimal disk (solid black line), superimposed to the models of \cite{Augereau2001} (blue dotted line) and \cite{Dent2014} (red dotted line). Note the clear mismatch in the location of the inner edge between this simulation and observational models.}
\label{fig:SimuSansD}
\end{figure*}

According to the theory proposed by \citet{Wisdom1980} and \citet{Mustill2012}, a planet can create a gap in a disk when its first-order mean-motion resonances $j:j\pm1$ overlap. This overlap is inevitable for a certain value of $j$, because as $j$ increases, the resonances are progressively located closer to the planet and closer to each other. \citet{Wisdom1980} provides an estimate for the critical value of $j$ at which this overlap occurs as
\begin{equation}
    j_\mathrm{overlap}=0.51\mu^{-2/7}\qquad,
\end{equation}
where $\mu$ is the ratio of planetary to stellar mass. The location of this critical resonance provides an estimate of the chaotic zone, and consequently, the size of the gap in terms of semi-major axis, given by
\begin{equation}
\left(\frac{\delta a}{a}\right)_\mathrm{chaos}=1.3\mu^{2/7}\qquad,
\label{da1}
\end{equation}
where the full gap size, including both the inner and outer regions relative to the planet's orbit, should be about twice this value. \citet{Mustill2012} highlight that this result depends on the planet's eccentricity. Equation~(\ref{da1}) remains valid for small eccentricities, up to a critical eccentricity estimated to
\begin{equation}
e_\mathrm{crit}=0.21\mu^{3/7}\qquad.    
\end{equation}
For larger planet eccentricities, a new regime applies where
\begin{equation}
\left(\frac{\delta a}{a}\right)_\mathrm{chaos}=1.8e^{1/5}\mu^{1/5}\qquad.
\label{da2}
\end{equation}

Let us assume that the inner edge of the  planetesimal disk is shaped by \bpb, as \bpc\ orbits much closer to the star. A numerical application using \bpb's parameters gives a critical eccentricity of $e_\mathrm{crit}\simeq0.02$. In this case, the large eccentricity regime described by Eq.~(\ref{da2}) applies, leading to a chaotic zone width of $(\delta a/a)_\mathrm{chaos}\simeq 0.4$. This places the outer edge of the gap at $\sim$14~au. This implies that the two known planets \bpb\ and \bpc\ alone lack the capacity to carve the disk up to 50~au, suggesting the presence of an additional planet.

However, this result should be interpreted with caution, as the formulae defined by \citet{Wisdom1980} and \citet{Mustill2012} are asymptotically valid for large $j$. Here, due to the high mass of \bpb, the overlap of the mean-motion resonances near \bpb\ occurs at $j_\mathrm{overlap} \simeq 2$, which is not particularly large. 

Another possible semi-analytical prediction that accounts for the high mass of \bpb\ is based on the work of \citet{Morrison2015}. These authors showed that the size of the outer chaotic zone, when the planet-to-star mass ratio $\mu$ exceeds $10^{-4}$, is described by
\begin{equation}
    \left(\frac{\delta a}{a}\right)_\mathrm{chaos} = 1.7 \mu^{0.31}\qquad.  
\end{equation}
In the case of \bpb, with $\mu=0.0065$, this formula predicts an outer gap edge at $\sim$13~au. This value is only slightly different than the $\sim$14~au obtained by the theory of \cite{Mustill2012}. However, it is important to note that the formalism of \citet{Morrison2015} does not account for the planet's eccentricity.

Thus, while existing semi-analytical approaches can accurately characterize the chaotic zone for planets that are either massive or eccentric, none of them simultaneously takes into account both regimes, nor consider small $j_\mathrm{overlap}$ values. Given these limitations, numerical simulations become indispensable for studying such regimes corresponding to the \bp\ case.

\begin{table}
\caption{Orbital solution for \bpb\ and \bpc\ used in our simulations.}
\label{tab:bcparameters}

\begin{tabular*}{\columnwidth}{@{\excs}lcc}

\noalign{\smallskip}\hline\hline\noalign{\smallskip}
& \bpb & \bpc \\
\noalign{\smallskip}\hline\noalign{\smallskip}
Mass $M$ [$\mjup$] & 11.90 & 8.89 \\
Semi-major axis $a$ [au] & 9.93 & 2.68 \\
Eccentricity $e$ & 0.103 & 0.32 \\
Inclination $i$ [\degr] & 89.00 & 88.95 \\
Arg. of periastron $\omega$ [\degr] & 199.3 & 66.0 \\
Long. of asc. node $\Omega$ [\degr] & 31.79 & 31.06 \\
Orbital phase $\tau$ & 0.719 & 0.724 \\
\noalign{\smallskip}\hline\noalign{\smallskip}
Stellar mass $M_*$ [$\msun$] & & 1.75\\
\noalign{\smallskip}\hline

\end{tabular*}
\tablefoot{These values follow the determination and the conventions of \citet{Lacour2021}. The reference epoch for the initial orbital phase $\tau$ is MJD 59 000 (May 31, 2020).}

\end{table}

\subsection{Simulation}

We present a simulation that includes the two known planets and a disk of planetesimals. The physical and orbital parameters of the planets are taken from the data set presented in \citet{Lacour2021} and are summarized in Table~\ref{tab:bcparameters}. The initial disk of planetesimals consisted of 400,000 massless particles that do not interact with each other. Their initial semi-major axes were randomly selected between 20 and 80~au, with initial eccentricities ranging from 0 to 0.05 and initial inclinations between 0 and $2\degr$ relative to the invariant plane of the two-planets system. The remaining initial orbital elements of the disk particles, including the longitudes of ascending nodes, arguments of periastron, and mean longitudes, were also randomly assigned values between 0 and $360\degr$. 

Calculations were performed using the Regularized Mixed Variable Step Size Symplectic (RMVS3) integrator \citep{Levison1994}, a modified version of the original Mixed Variable Symplectic (MVS) scheme by \citet{Wisdom1991}, which includes a first-order but rapid treatment of close encounters. This approach is particularly relevant here, as we aim to examine the location of the disk's inner edge as sculpted by the planets, specifically focusing on particles near the instability region. The integration was carried out over 20~Myr, corresponding to the adopted age of the system. 

The results are presented in Fig.~\ref{fig:SimuSansD}. As shown in the figure, the planets carve the disk out to  $\sim$28~au. This is inconsistent with observations that suggest an inner disk cavity extending out to 50~au. There are still far too many particles remaining between  $\sim$28 and 50~au. Additional tests, varying the orbits of the planets within the uncertainties provided by \citet{Lacour2021}, did not alter this outcome. 

Given this result, we can therefore conclude that the semi-analytical theories of \citet{Mustill2012} and \citet{Morrison2015} seem to underestimate the range of gravitational interactions of \bpb, which is either too massive or too eccentric. However, the conclusion remains the same: the inner edge of the planetary disk at 50~au cannot be attributed solely to the perturbative action of \bpb\ and \bpc. In the following sections, we explore how this could be achieved with the introduction of a hypothetical additional planet orbiting outside \bpb's orbit.

\section{An additional planet}
\label{1AdditionalPlanet}
We now hypothesize the presence of an additional planet in the \bp\ system, which we shall refer to as \bpd. We present simulations using the same initial  planetesimal disk as before, but this time assuming a three-planets system. The initial orbital parameters of \bpb\ and \bpc\ were still set to the values listed in Table~\ref{tab:bcparameters}.

\subsection{Constraints and initial orbital parameters}
\label{Constraints}

Since this planet is proposed to be part of the same planetary system as the two known planets, it is reasonable to assume that it is more or less coplanar with them. Therefore, we assumed that its inclination relative to the midplane of the two-planet system does not exceed $2\degr$. Angular parameters such as the longitude of the ascending node, the longitude of periastron, and the initial mean longitude were selected randomly, as these are expected to secularly process under the gravitational influence of \bpb\ and \bpc\ and thus play a minor role in the long term.

The most critical parameters to determine for this hypothetical \bpd\ are its mass $m_d$, orbital semi-major axis $a_d$, and eccentricity $e_d$. Several constraints must be satisfied for each of these parameters.

The first constraint is observational. Current observations, such as those made with Spectro Polarimetric High contrast Exoplanet REsearch (SPHERE), rule out the presence of super-Jupiter planets exterior to \bpb\ \citep{Lagrange2020}.

Second, based on the semi-analytical theories discussed in Sect~\ref{Theory}, reproducing a disk with an inner edge at 50~au requires \bpd\ to be located at a semi-major axis no greater than $\sim$40~au. Although these semi-analytical theories are not always applicable, this constraint provides an indicative range for the planet's semi-major axis.

\begin{figure}
    \centering
    \includegraphics[width=\columnwidth]{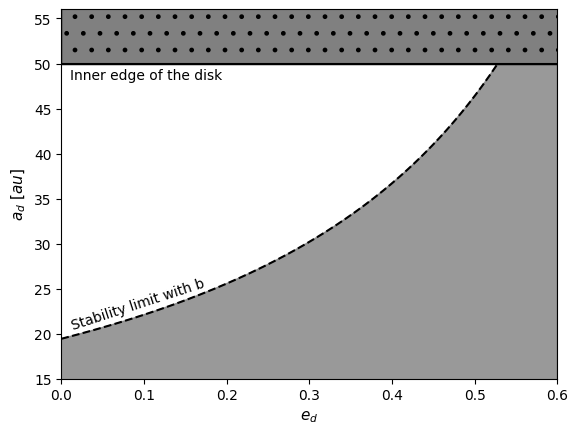}
    \caption{Accessible semi-major axis of the hypothetical additional planet \bpd, prior to simulations, as a function of its eccentricity. Shaded areas are not accessible by \bpd. Its semi-major axis must be smaller than the inner edge of the disk at 50~au (solid black line) to sculpt it up to this distance, while being greater than the stability limit with \bpb\ (dashed black line) to avoid close encounters, based on the criterion of \cite{Petrovich2015}.}
    \label{fig:Bpic-d-constraint-ea}
\end{figure}

Finally, the planet must remain dynamically stable against perturbations from other planets. Assuming that \bpc\ is too close to the star, we focused solely on the two-planet system comprising \bpb\ and \bpd. According to \cite{Petrovich2015}, a stability criterion between these two planets, with mass ratios relative to the central star such that $10^{-2} \geq \mu_b > \mu_d \geq 10^{-4}$, can be defined by
\begin{equation}
    \frac{a_{d}(1-e_{d})}{a_{b}(1+e_{b})} > 2.4 \, {\mu_{b}}^{\frac{1}{3}} \left(\frac{a_{d}}{a_{b}}\right)^{\frac{1}{2}} + 1.15
    \qquad.
\end{equation}
This stability condition between \bpb\ and \bpd\ directly imposes a lower limit on \bpd's semi-major axis as a function of its eccentricity. This relationship is depicted in Fig.~\ref{fig:Bpic-d-constraint-ea}. As the eccentricity increases, the permissible mass range for \bpd, constrained by this stability condition with \bpb\ and the inner edge of the disk, narrows progressively. For eccentricities exceeding $\sim$0.5, no viable solutions remain.

Therefore, our study focused on a planet \bpd\ with a mass $m_d\leq$ 1~\mjup, a semi-major axis $a_d$ between $\sim$30 and $\sim$40~au, and an eccentricity $e_d\la$ 0.5. For the eccentricity, we first assumed a moderate value in line with \bpb\ and initially set it at $e_d=$ 0.05.

\subsection{Results}

\begin{figure*}
\makebox[\textwidth]{
\includegraphics[width=0.3333\textwidth]{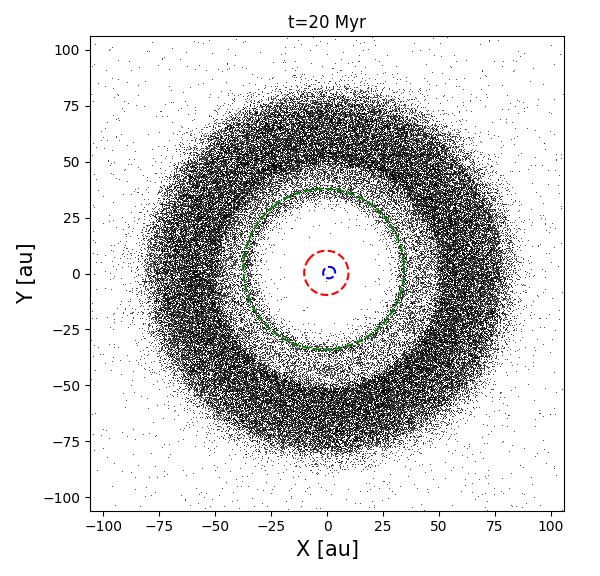} \hfil
\includegraphics[width=0.3333\textwidth]{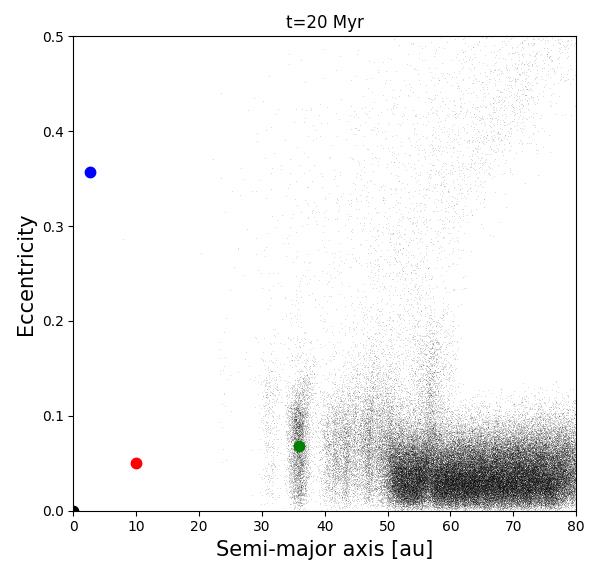} \hfil
\includegraphics[width=0.3333\textwidth]{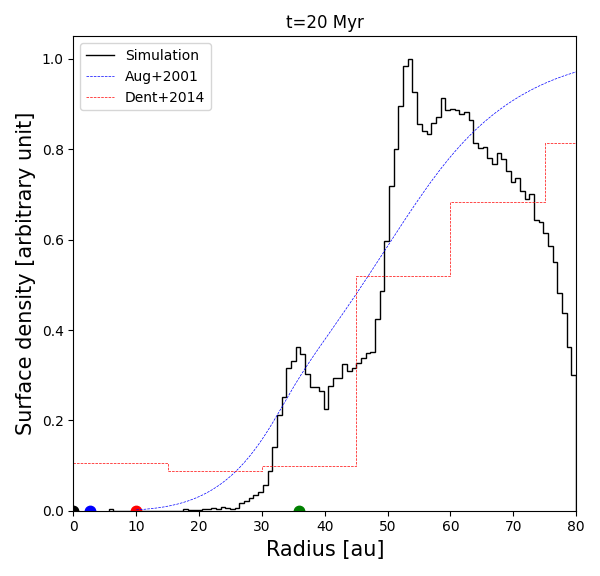}}
\caption{Example of a simulation of the dynamics of the \bp\ planetary system with three planets, \bpb\ (red), \bpc ~(blue), and an additional planet \bpd\ (green), and the disk of planetesimals (black). The initial orbital parameters of \bpb\ and c are taken from Table~\ref{tab:bcparameters}. In this example, the initial orbital parameters of \bpd\ are: $m_d=0.2$~\mjup, $a_d=35$~au, and $e_d=0.05$. The plotting conventions are identical to Fig.~\ref{fig:SimuSansD}.}
\label{fig:ExSimuValide}
\end{figure*}
Figure~\ref{fig:ExSimuValide} illustrates an example of a simulation with three planets that successfully reproduces the desired disk profile. As opposed to the previous simulation, we now find that the inner cavity aligns with the observational models. 

\begin{figure}
\centering
\includegraphics[width=\columnwidth]{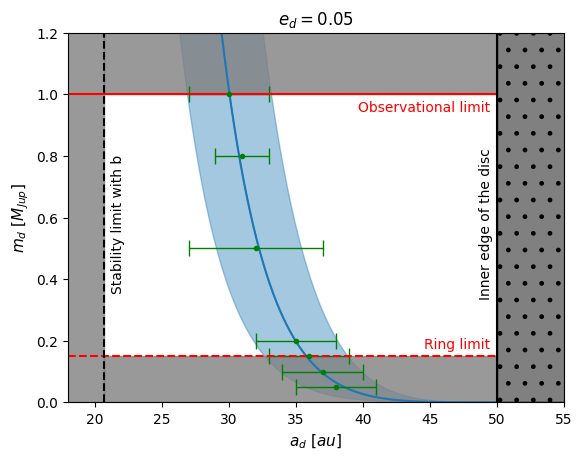}
\caption{Combinations of \bpd's mass and semi-major axis within observational constraints that successfully reproduce the disk profile at 50~au. For several values of $m_d$, the corresponding acceptable ranges of $a_d$ are displayed in green. A power-law fit, according to Eq.~(\ref{PowerLaw}), is overlaid in blue, taking into account the ranges of the semi-major axis. Shaded areas are not accessible by \bpd. The limits in semi-major axis are similarly presented as in Fig.~\ref{fig:Bpic-d-constraint-ea}. Additionally, above a mass of 1~\mjup, corresponding to the observational limit (solid red line), \bpd\ would have already been detected in previous observations (\cite{Lagrange2020}). And below 0.15~\mjup, corresponding to the ring limit (dashed red line), \bpd\ does not completely clear the inner zone.}
\label{fig:Bpic-d-constraint-am}
\end{figure}

Of course, the solution is not unique. If we assume a more massive \bpd, it would likely create a larger gap in the disk, necessitating its placement closer to the star to reproduce a disk edge at 50~au. For each assumed value of $m_d$, we should be able to identify a corresponding $a_d$ that yields the desired inner edge of the disk. To explore suitable combinations of $m_d$ and $a_d$, that can reproduce the desired disk profile, we conducted several dozen simulations. The overall results of this investigation are summarized in Fig.~\ref{fig:Bpic-d-constraint-am}, which displays the acceptable ranges of $a_d$ for various values of $m_d$. As $a_d$ depends, to within one translation, on the position of the simulated disk inner edge at 50~au, the acceptable range for $a_d$ corresponds to the measurement uncertainty of this simulated inner edge. A wide acceptable range of semi-major axes therefore indicates a very unclear inner edge. 

In addition, Fig.~\ref{fig:Bpic-d-constraint-am} includes an attempt to adapt to a power law, similar to that of the semi-analytic theories discussed in section \ref{Theory}, according to
\begin{equation}
    \frac{\delta a_d}{a_d} = c_1 {\mu_d}^{c_2}
    \quad \Leftrightarrow \quad
    m_d = \left(\frac{\delta a_d}{c_1 a_d} \right)^{\frac{1}{c_2}} M_* \qquad.
    \label{PowerLaw}
\end{equation}
To achieve an inner edge of the disk at 50~au, we set
\begin{equation*}
    \delta a_d = 50-a_d \qquad [\text{au}],
\end{equation*}
where $c_1$ and $c_2$ are free parameters. For a low eccentricity of 0.05, the resulting $c_1\simeq5.1$ is significantly higher than that predicted by the theories discussed in Sect.~\ref{Theory}. Here again it appears that these theories underestimate the amplitude of the chaotic zone of \bpd\ because its value of $j_\mathrm{overlap}\simeq7$ is still relatively small. However, the phenomenon behavior indicated by $c_2\simeq0.27$ is reminiscent of the predictions of 0.29 and 0.31 from the theories of \citet{Mustill2012} and \citet{Morrison2015}, respectively.

As expected, a more massive planet must be positioned closer to the star. This imposes an independent upper limit on $m_d$. For $m_d$ exceeding a certain value, \bp\ would need to be placed too close to the stability threshold with \bpb, making it impossible to achieve a stable and suitable configuration. Such values of $m_d$ are already excluded here by the observational constraints established by \citet{Lagrange2020}. Nonetheless, this consideration remains crucial for the subsequent analysis.

\begin{figure*}
\makebox[\textwidth]{
\includegraphics[width=0.3333\textwidth]{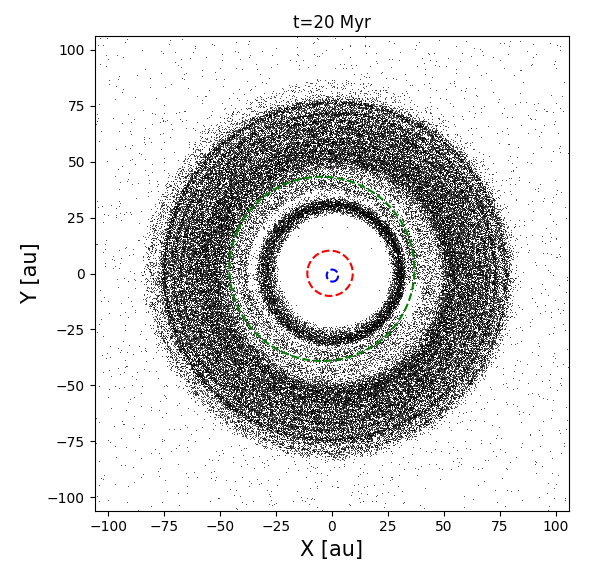} \hfil
\includegraphics[width=0.3333\textwidth]{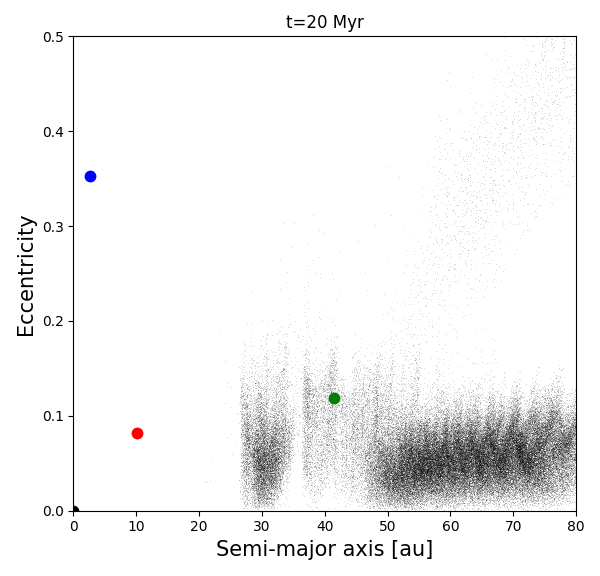} \hfil
\includegraphics[width=0.3333\textwidth]{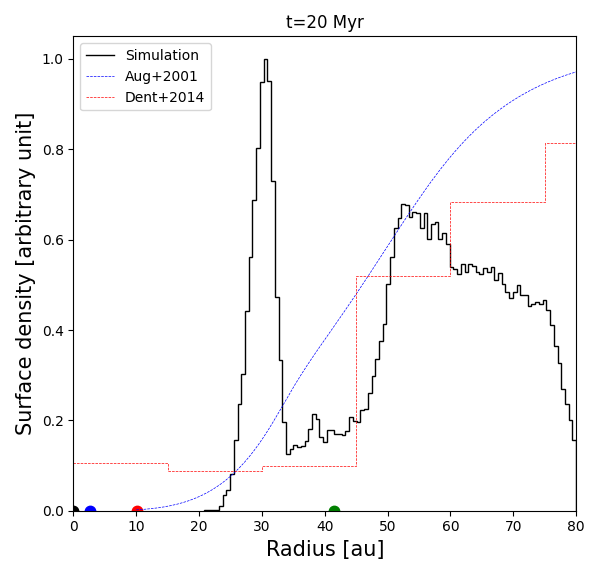}}
\caption{Example of a simulation of the dynamics of the \bp\ planetary system with three planets, \bpb\ (red), \bpc\ (blue) and an additional planet \bpd\ (green), and the disk of planetesimals (black). The initial orbital parameters of \bpb\ and c are taken from Table~\ref{tab:bcparameters}. In this example, the initial orbital parameters of \bpd\ are: $m_d=0.05$~\mjup, $a_d=40$~au, and $e_d=0.05$. The plotting conventions are identical to Fig.~\ref{fig:SimuSansD}. This combination still generates a disk inner edge at 50~au, but there is now enough space between \bpb\ and \bpd\ to allow an additional ring of particles to exist at $\sim$30~au.}
\label{fig:ExSimuInvalide}
\end{figure*}
Our study also allows us to derive a lower limit for $m_{d}$. A low-mass \bpd\ generates a small gap in the disk and must therefore be positioned closer to 50~au to effectively carve the disk to this distance. However, this simultaneously creates a wider region between \bpb\ and \bpd\ where planetesimals could potentially orbit safely. This is illustrated in Fig.~\ref{fig:ExSimuInvalide}, which shows the output of a simulation similar to that of Fig.~\ref{fig:ExSimuValide}, but with $m_{d}\la0.15$~\mjup. In agreement with the overall results shown in Fig.~\ref{fig:Bpic-d-constraint-am}, this combination successfully reproduces the inner edge of the disk at 50~au. However, it leaves enough space between \bpb\ and \bpd\ for an additional ring of stable particles to exist around $\sim$30~au. We indeed recover the stability limit of particles at  $\sim$28~au due to \bpb, as \bpd\ is now too far away and not massive enough to destabilize the planetesimals orbiting in this region. This additional ring of particles is, of course, incompatible with the observational models of \citet{Augereau2001}, \citet{Dent2014}, and \citet{Ballering2016}. Therefore, this configuration must be excluded.

\section{Alternative scenarios}
\label{AlternativeScenarios}
The solutions presented in Sect.~\ref{1AdditionalPlanet}, which successfully reproduce the desired inner edge of the planetesimal disk, rely on specific initial parameter choices. But, other combinations of initial parameters might also yield acceptable solutions. In Fig.~\ref{fig:ExSimuInvalide}, we demonstrate that a low eccentricity and low-mass additional planet is insufficient to efficiently carve the disk up to 50~au, resulting in a stable, unperturbed ring of planetesimals around $\sim$30~au, which is inconsistent with observational models. This observation led us to derive a minimum mass of  $\sim$0.15~\mjup\ for \bpd. This result is directly linked to our consideration of only one additional planet with low eccentricity. To address this limitation, in Subsection \ref{EccentricPlanet}, we explore the possibility that the additional planet could have a higher eccentricity than initially assumed. Additionally, in Subsection \ref{SeveralPlanets}, we consider the potential presence of two additional planets instead of just one.

\subsection{A more eccentric additional planet}
\label{EccentricPlanet}
The simulations presented in Sect.~\ref{1AdditionalPlanet} all assumed an initial low eccentricity of 0.05 for the additional planet. Although this eccentricity undergoes secular evolution due to gravitational perturbations from \bpb\ and \bpc, the simulations indicate that \bpd's eccentricity varies only slightly, never exceeding $\sim$0.1. It is therefore of interest to study the potential for larger initial eccentricities of \bpd.

A preliminary estimate of the outcome for such a configuration can be obtained by requiring that the apoastrons of \bpd\ coincide in both scenarios. We anticipated that an eccentric planet would sculpt the disk in the same way as a low eccentricity planet with the same apoastron. Since eccentric planets dig the disk inwards and outwards more efficiently than a non-eccentric planet, this hypothesis suggests that they would have to move closer to the star to maintain a similar disk structure. 

\begin{figure*}
    \makebox[\textwidth]{
    \includegraphics[width=0.3333\textwidth]{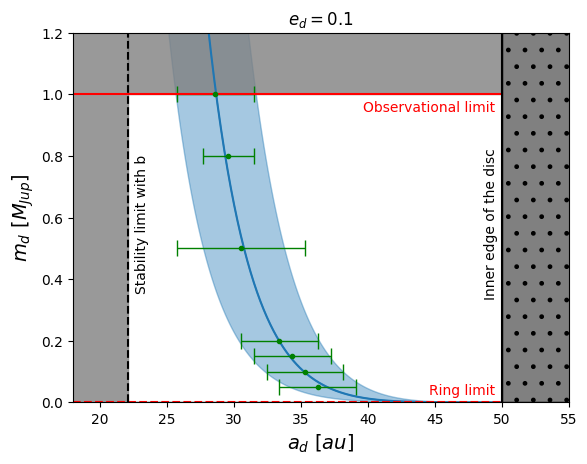} 
    \includegraphics[width=0.3333\textwidth]{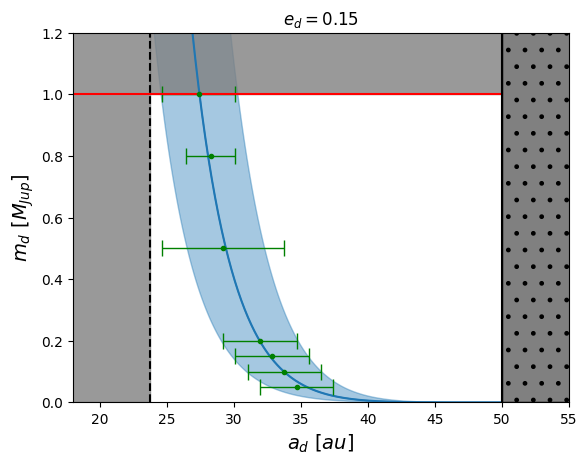}
    \includegraphics[width=0.3333\textwidth]{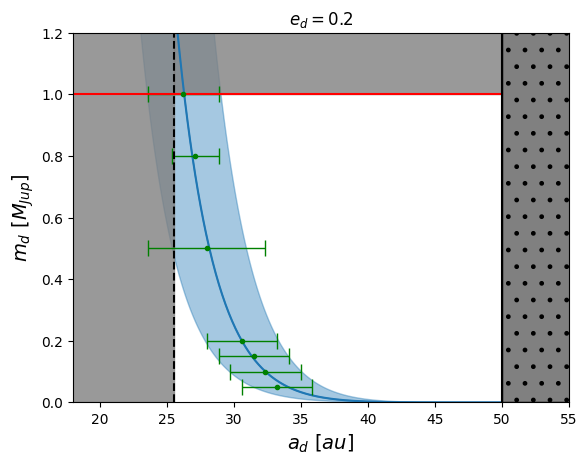}}
    \makebox[\textwidth]{
    \includegraphics[width=0.3333\textwidth]{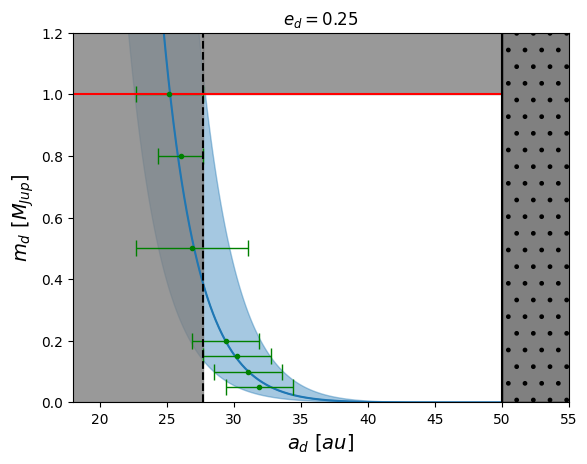}
    \includegraphics[width=0.3333\textwidth]{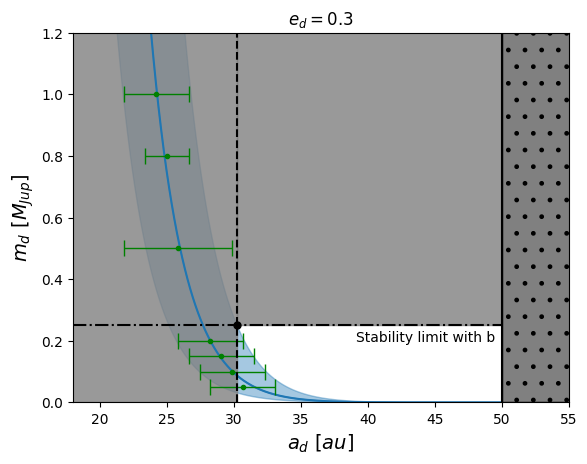}
    \includegraphics[width=0.3333\textwidth]{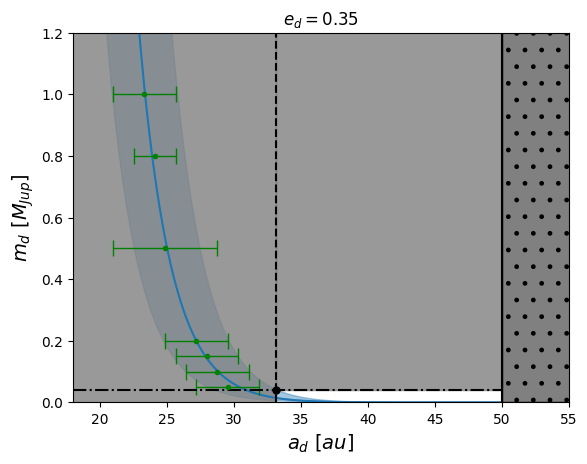}}
    \caption{Combinations of \bpd’s mass and semi-major axis for various eccentricities, within observational constraints that successfully reproduce the disk profile at 50~au. The plotting conventions are identical to Fig.~\ref{fig:Bpic-d-constraint-am}. The ring limit quickly tends to zero as the eccentricity increases. The minimum semi-major axis stability limit with \bpb\ imposes a maximum mass limit (dot-dashed black line), determined by the intersection (black dot) with the fit's right extremum of the acceptable range of \bpd's semi-major axis. This maximum mass limit decreases as eccentricity increases, becoming more restrictive than the observational mass limit for $e_d \protect \ga 0.25$, and tends towards zero as eccentricity continues to increase (see Fig.~\ref{fig:Bpic-d-constraint-em}).}
    \label{fig:Eccentricities}
\end{figure*}

We numerically explored various configurations with increasing initial values of \bpd's eccentricity, similarly to the approach presented in Sect.~\ref{1AdditionalPlanet}. The results are summarized in Fig.~\ref{fig:Eccentricities}, which presents plots similar to the one in Fig.~\ref{fig:Bpic-d-constraint-am} for the increasing eccentricity regimes of \bpd. For each eccentricity, we again fitted the semi-major axis by a power law according to Eq.~\ref{PowerLaw}. 

As eccentricity increases, the phenomenon evolves, as shown by the corresponding values of $c_2$ in the Table~\ref{tab:FitParameters}. It remains consistent with the predictions of \cite{Mustill2012} but diverges from those of \cite{Morrison2015}. This is expected, as the theory of \cite{Mustill2012} accounts for high eccentricity of $\sim$0.3, whereas \cite{Morrison2015} does not. However, for extreme eccentricities of $\sim$0.5, our values no longer align with either theory, as they do not apply to such eccentricities.

\begin{figure}
    \centering
    \includegraphics[width=\columnwidth]{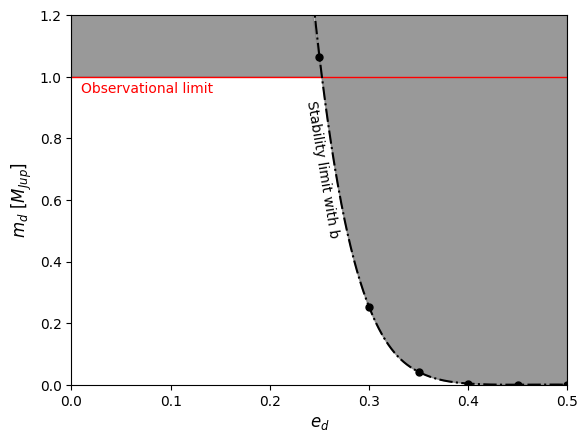}
    \caption{Accessible mass of the potential \bpd\ as a function of its eccentricity $e_{d}$, based on the results of Fig.~\ref{fig:Eccentricities}, whose plot conventions are similar.}
    \label{fig:Bpic-d-constraint-em}
\end{figure}

As expected, increasing eccentricity effectively clears the remaining dust ring when present, rapidly lowering the lower mass limit and extending the range of acceptable parameters to smaller masses. However, with higher eccentricity, \bpd\ becomes more likely to enter the region of close encounters with \bpb, falling below the stability limit determined in Sect.~\ref{Constraints}. The higher the eccentricity, the smaller the range of acceptable masses for \bpd. Thus, for each eccentricity regime, a maximum mass is defined, as summarized in Fig.~\ref{fig:Bpic-d-constraint-em}. As eccentricity increases, this maximum mass limit decreases, becoming more restrictive than the observed mass limit for $e_d \ga$ 0.25, and tends towards zero as eccentricity continues to increase. Beyond an eccentricity threshold of $\sim$0.4, it becomes impossible for a planet, regardless of its mass and semi-major axis, to match the observational constraints while remaining stable within the system.

\begin{table}
    \centering
    \caption{Values of the fit parameters, according to Eq.~(\ref{PowerLaw}), for each eccentricities tested.}
    \begin{tabular*}{\columnwidth}{@{\excs}l cccccccccc}
        \noalign{\smallskip}\hline\hline\noalign{\smallskip}
        \textbf{$e_d$} \hspace{0.25em} & 0.05 & 0.10 & 0.15 & 0.20 & 0.25 & 0.30 & 0.35 & 0.40 & 0.45 & 0.50 \B \\ \hline
        \textbf{$c_1$} \hspace{0.25em} & 5.1 & 4.8 & 4.7 & 4.7 & 4.7 & 4.7 & 4.8 & 4.8 & 4.9 & 5.0 \T \\
        \textbf{$c_2$} \hspace{0.25em} & 0.27 & 0.25 & 0.23 & 0.22 & 0.21 & 0.20 & 0.19 & 0.18 & 0.18 & 0.17 \B \\ \hline
    \end{tabular*}
    \label{tab:FitParameters}
\end{table}

\subsection{Several additional planets}
\label{SeveralPlanets}
\begin{figure*}
\makebox[\textwidth]{
\includegraphics[width=0.33\textwidth]{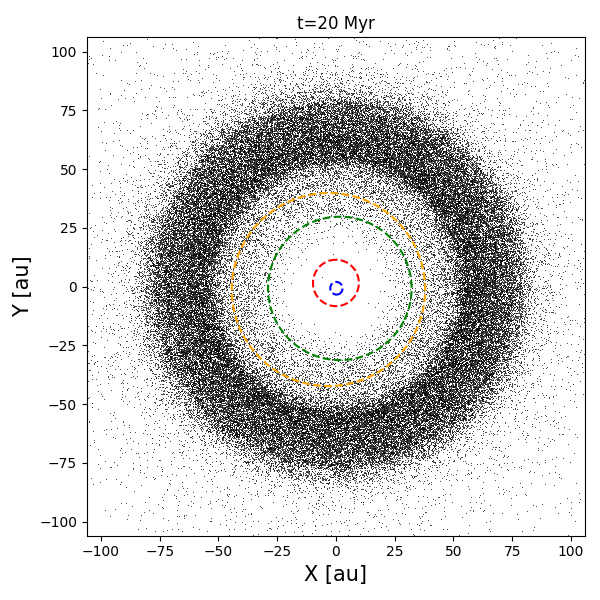} \hfil
\includegraphics[width=0.33\textwidth]{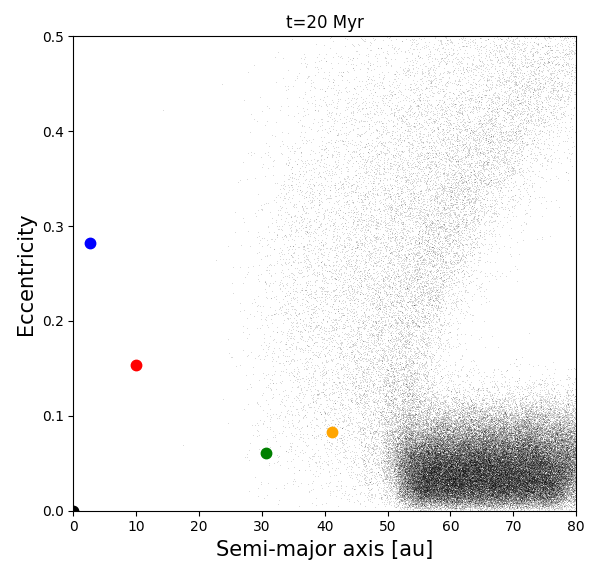} \hfil
\includegraphics[width=0.33\textwidth]{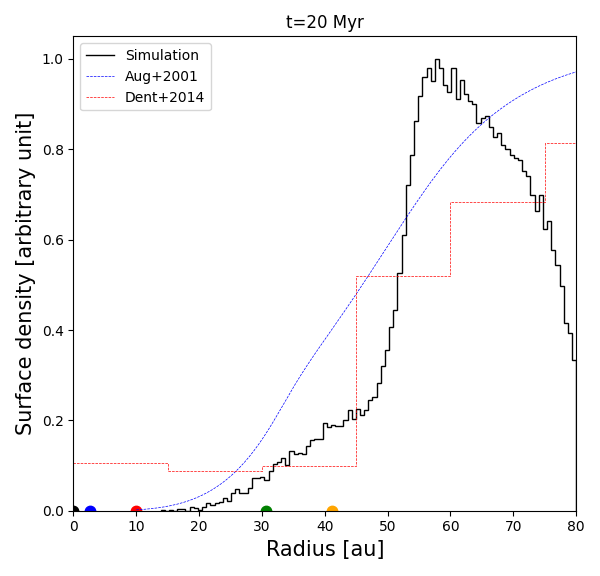}}
\caption{Example of a simulation of the dynamics of the \bp\ planetary system with four planets, \bpb\ (red), \bpc\ (blue) and two additional planets (green and orange), and the disk of planetesimals (black). The initial orbital parameters of \bpb\ and c are taken from Table~\ref{tab:bcparameters}. In this example, the additional planets have a mass of 0.05~\mjup, an initial eccentricity of 0.05, and initial semi-major axes of 30 and 40~au respectively. The plotting conventions are identical to Fig.~\ref{fig:SimuSansD}.}
\label{fig:ExValid2planet}
\end{figure*}

In this section, we briefly explore the possibility of having two additional planets in the \bp's planetary system instead of one. The available parameter space for such a configuration is significantly limited. Clearly, these two planets must remain dynamically stable. Any configurations where a single additional planet is sufficient to efficiently clear the disk up to 50~au should not be considered here, as the insertion of an other additional planet would inevitably lead to instability. We focused on configurations like those illustrated in Fig.~\ref{fig:ExSimuInvalide}, where a single additional planet is insufficient. Furthermore, Figs.~\ref{fig:Eccentricities}~and~\ref{fig:Bpic-d-constraint-em} show that as the eccentricity of the additional planet increases, the space available for any other additional planets is greatly reduced. Therefore, we worked on a model involving two additional planets, both with low mass ($\leq$0.15~\mjup) and a low eccentricity of $\sim$0.05.

Figure~\ref{fig:ExValid2planet} illustrates a simulation involving two additional planets which successfully carves the disk up to 50~au without leaving a stable ring at $\sim$30~au. We explored various similar configurations by varying the masses and locations of the planets. Considering all requirements, the available parameter space around the configuration shown in Fig.~\ref{fig:ExValid2planet} is quite limited. Notably, it appeared that among the two additional planets, the inner one must have a mass at least equal to that of the outer one to ensure its dynamical stability.

The existence of two additional low-mass, low-eccentricity planets presents therefore an interesting alternative to the model of a single, more massive planet. However, this hypothesis was not explored further, as these planets are relatively small and well below current detection limits, making it unnecessary to immediately pursue this avenue.
\section{Conclusion and discussion}
\label{Conclusion}
This paper investigates how the \bp\ planetary system can dynamically clear out the disk of planetesimals up to the $\sim$50~au threshold, as modeled by \citet{Augereau2001}, \citet{Dent2014}, and \citet{Ballering2016}. This analysis was conducted under the assumption that the planetesimal disk initially extends from the star to well beyond 80~au, without accounting for potential planetary migration or other dynamical effects.

The first result is that the currently known planetary system is unable to clear the disk as needed. With only \bpb\ and \bpc, the disk would not be carved beyond $\sim$28~au. Therefore, the presence of additional, yet undiscovered planets, such as \bpd, at greater distances can be hypothesized. 

Our simulations indicate that an additional planet is sufficient to achieve the desired outcome. Various combinations of mass, semi-major axis, and eccentricity are possible, but the higher the eccentricity, the more restricted the parameter space becomes. In any case, the planet’s eccentricity cannot exceed $\sim$0.4. Alternatively, we show that a model involving two planets of low mass and low eccentricity, instead of a more massive one, could also be a viable option.

However, it is important to note that our dynamical analysis could be further refined. We did not attempt to reproduce the complete radial profile of the disk, and simplified the problem to a single radius representative of the inner edge, which is itself uncertain due to the difficulty of deprojecting the observations. As some of the simulated profiles have a complex structure, this approximation introduces uncertainties about the semi-major axis of the additional planets. 

Also, we did not explore specific dynamical configurations, such as a planet trapped in an outer mean-motion resonance with \bpb. In such a case, a phase-protection mechanism could prevent \bpd\ from being ejected while still orbiting within the instability zone of \bpb, similar to the relationship between Neptune and Pluto in the Solar System \citep{Greenberg1977}. Future modeling efforts should focus on these specific scenarios, but we emphasize that a more accurate understanding of the exact orbital parameters of \bpc\ and \bpb\ is necessary first. Indeed, \citet{Beust2024} demonstrated that these two planets could be temporarily locked in a 7:1 mean-motion resonance. If this were the case, any additional planets in resonance with \bpb\ would form a chain of resonances involving three planets, as discussed for example in the case of HR8799 \citep{Wang2018}. This could lead to a significantly different dynamical result.

As it stands, the observed disk structure could therefore be explained by the presence of at least one additional planet, with a mass still well below the current detection limit for high-contrast imaging, estimated at $\sim$1~\mjup\ \citep{Lagrange2020}. Recent observations with JWST \citep{Krammerer2024} could potentially improve upon this limit. However, as this depends on the projected separation between the planet and the star, rather than the semi-major axis of the orbit, it is difficult to draw definitive conclusions about the presence of a planet. In the most favorable scenario, a planet with low eccentricity, and an apparent separation equal to its semi-major axis of 35~au at the time of observation, should be detectable if it has a mass of at least 0.3~\mjup\ (the equivalent of Saturn). No planet of this type has been observed, but it is possible that the planet is in an unfavourable configuration, perhaps occulted by \bp, or that it is completely absent. So far, the available JWST data do not yet allow us to reduce the detection limit below the current limit of $\sim$1~\mjup\ \citep{Lagrange2020}. We hope that future observations will enable us to lower this limit and strengthen the constraints on our model, or even detect the suspected planet directly.

It would now be interesting to examine the impact of a system with three or even four planets on the disk, in order to try to understand the dynamical origin of the various intriguing features observed so far, particularly the asymmetries \citep{Kalas-Jewitt1995, Heap2000} and the clump \citep{Telesco2005, Dent2014, Matra2017}, through gravitational interactions with the planets.
\begin{acknowledgements}
\newline We are grateful to an anonymous referee for feedback that helped us improve this manuscript. All computations presented in this paper were performed using the GRICAD infrastructure (\texttt{https://gricad.univ-grenoble-alpes.fr}), which is supported by Grenoble research communities. 
This project has received funding from the European Research Council (ERC) under the European Union's Horizon 2020 research and innovation program (COBREX; grant agreement n$^\circ$885593).
V. Faramaz acknowledges funding from the National Aeronautics and Space Administration through the Exoplanet Research Program under Grants No. 80NSSC21K0394 (PI: S. Ertel) and No. 80NSSC23K0288 (PI: V. Faramaz). 
\end{acknowledgements}

%
\bibliographystyle{aa.bst}
\bibliography{bibli}
\end{document}